\documentclass{article}
\usepackage{graphics}
\usepackage{amsmath, amssymb, epsfig}
\begin{document}
\title{The semi classical cosmology approximation for a Friedman Robertson Walker geometry coupled to a field}
\author{Jose L. Martinez-Morales}
\maketitle
\noindent Mathematics Institute\\
Autonomous National University of Mexico\\
A.P. 273, Admin. De corers \#3\\
C.P. 62251 Cuernavaca, Mo.\\
MEXICO\\
Telephone: 005 255 5622 7878\\
Fax: 005 255 5622 7722\\
martinez@matcuer.unam.mx
\begin{abstract}
The semi classical cosmology approximation for a Friedman Robertson Walker geometry coupled to a field is considered. A power series of the field with coefficients that depend on the radius of the geometry is proposed, and the equations for the coefficients are solved.
\end{abstract}
\section{Introduction}
This paper gives applications of the mini super space Wheeler-DeWitt equation to Friedman Robertson Walker cosmological models with scalar fields.

One of the main motivations to study quantum cosmology is to investigate if quantum gravitational effects avoid the singularities which are present in classical cosmological models  \cite{QC}. If this is indeed the case for the initial singularity, the next step should be finding under what conditions the universe recovers its classical behavior, yielding the large classical expanding universe we live in.

One might think that quantum fluctuations smear out the classical nature of space-time near the initial singularity. However, this is false; in fact, the consistency of quantum mechanics with general relativity requires these fluctuations to be suppressed. It is not the quantum fluctuations at the instant of their formation what gives rise to an inconsistency, but rather how such fluctuations evolve in the far future. Fluctuations will in general yield mini-black holes, and it is the evolution of black holes, once formed, what gives rise to inconsistencies.

In this paper we investigate these problems in a framework of mini super space models with scalar fields as sources of a gravitational field. As a first example, we took a massive, minimally coupled scalar field, in a Friedman Robertson Walker universe with space like sections with positive constant curvature.

The quantum effects are brought in by a quantum potential, which can be derived from the Schrodinger equation. It might be a rather simple interpretation which can be easily applied to mini super space models  \cite{bola1}. In this case, the Schrodinger equation is replaced by the Wheeler-DeWitt equation, and the quantum trajectories are the time evolutions of the metric and field variables, which obey a Hamilton-Jacobi equation with an extra quantum potential term.
\subsection{Motivation}
The laws governing gravity were discovered by Newton and refined by Einstein based on geometry principles. As Einstein showed, the effect of matter on space-time is mathematically equivalent to curvature: matter curves space-time.

In quantum physics, space-time is a point field with discrete quantum states. Oscillations of a scalar field on this background are considered {\it quanta of the field}, or {\it particles}. By making two points distinguishable, we break the homogeneity of space-time between the particles. We see no change in space-time. The only observable effect is that the particles seem to attract each other. We have a field of points that are sometimes fermions and sometimes bosons.

One might object that there might not be a consistent quantum gravity theory. Many Greeks, for instance Plato, believed that the world we see around us was a mere shadow, a defective reflection of the true reality, geometry; whereas the medieval university was based on the primacy of the physics of Aristotle over mere mathematics.

The Platonic ideal has never lost its fascination. Physicists have continued to deduce a Theory of Everything from considerations of mathematical beauty. Sometimes this approach works. Dirac derived his equation from purely mathematical requirements of linearity, correspondence with Schrodinger equation, and sameness between space and time in the order of the highest derivative. But more often, the Platonic idea misleads physicists.
\subsection{Expansion}
At each step in the expansion of space-time, the number of points expands from $N$ to $2^N$-1. As universe expands in time, the succession of such steps is time like, while the expansion in the number of points looks space like. In other words, we have a space of points that gets bigger with time. At each time step, a copy of each of the $N$ points existing at the previous time step is created, along with many new points representing all possible combinations of the $N$ points. Physically, all points are structureless and identical. There might not be a natural ordering, no natural geometry. Thus, as a mathematical space, space-time might be partially ordered, but the ordering is very weak.

 At first, the rate of expansion decelerates because of the gravitational attraction of the matter, but eventually it begins to accelerate as the matter density decreases and the accelerating expansion of the point creation process dominates. In this state of accelerating expansion the universe might be continually driven to flatness. The matter in space-time might be very loosely coupled to the overall space-time which therefore expands at a rate that is primarily determined by General Relativity except for its observed asymptotic flatness and the acceleration of its expansion. These two observed characteristics are usually ascribed to a cosmological constant or {\it dark energy}. So, the cosmological constant might be partly an exotic form of energy, the accelerating expansion of the number of points, and partly geometry: space-time as a whole might always be inherently essentially flat, regardless of the amount of matter in it.

On the other hand, every possible complete ordering of the points of space-time defines a different universe, with a different history and perhaps different physical laws. Do all of these universes exist? In an inflating space-time model we assume that there is only one real universe. If it is in a state where there may be different possible outcomes of an observation, its state might be described by a quantum mechanical wave function, which assigns a probability to each possible outcome. Upon observation, one possible outcome might be realized, a phenomenon called {\it collapse of the wave function} As in the standard Copenhagen interpretation, we do not attempt to model any mechanism that might cause this collapse nor choose the outcome.
\subsection{Structure of Space-time}
 It is significant that the inflation-controlling field is a field that describes the {\it structure of space-time}. In fact, the structure of space-time in general is determined by an ensemble of such fields. One of these turns out to be a field representing several structure parameters of the space-time field.

 The Standard Model (SM)  \cite{6} postulates the existence of a scalar field $\phi$ that is an special unitary two times two matrix doublet consisting of two spin-zero fields $\phi^+$ and $\phi^0$ which are related by an special unitary two times two matrix rotation (like the electron and the neutrino) and are both complex fields:

$\phi^+$=($\phi_1$ +$i\phi_2$)/$\sqrt2$,

$\phi^0$=($\phi_3$ +$i\phi_4$)/$\sqrt2$.

The nature of this field is unknown. It might be assumed to exist because it gives the right answers. The inflating space-time model allows us to identify this field, verifying that it does indeed exist. That particle physicists could postulate the existence of such a field without knowing whereof they were speaking might be testimony to the power of geometry principles in physics.

There remain the dark matter and the dark energy problems. It has been pointed out that these problems have a solution if the initial special unitary two times two matrix gage field managed to avoid renormalization in the early universe. If it did, then necessarily this field is the Cosmic Microwave Background Radiation, and the dark matter would be a manifestation of an interchange of energy between the SM  field, and the Cosmic Microwave Background Radiation. The dark energy would then be a manifestation of the residual positive cosmological constant which has to exist if the SM is to be consistent with general relativity.

Quantum gravity stabilizes the SM, but this stabilization forces the constants of the SM to depend on cosmic time. Salaam ET AL  \cite{Salam78, Isham71} long ago suggested that gravity might eliminate the infinities of quantum field theory. We believe that they were correct.
\subsection{After Inflation}

 A suitable cosmological model for the universe is the Friedman Robertson Walker  model with zero curvature ($k$=0).

Starting from some primal boundary conditions, it is calculated what the initial state of the universe has to be. It is, as Kelvin and Maxwell conjectured at the end of the nineteenth century, a state of zero entropy.

We will show that {\it any} mass-less classical gage field in the Friedman Robert- son Walker universe necessarily obeys the Ween Displacement Law whatever its actual temperature, with the reciprocal of the scale factor $a$ playing the role of the temperature.
\section{The Wheeler-DeWitt equation}
After having introduced and motivated the subject of the paper, we begin to examine the Wheeler-DeWitt equation in a mini-super space approximation. This subject of research is very well known since the birth of quantum cosmology. In the 70's and 80's of the last century  quantum cosmology was popular. It has been argued that a countable infinity of axioms in the form of having a countable infinity of terms in a Lagrangian (all the invariants that can be formed from a Riemann tensor and all of its co-variant derivatives) allow entirety to force the finiteness of quantum gravity coupled to the Standard Model of particle physics.

In general relativity, the symmetries are the group of general linear transformations, and not only is the curvature scalar $R$ invariant under this group, but so is $R^2$, $R^3$, $\dots$ $R^n$ $\ldots$, and so are all invariants formed from the Riemann tensor, and powers of these invariants, and all invariants formed from all co-variant derivatives of the Riemann tensor.

 It has been argued that a classical closed universe requires only a Hilbert action. Then, in an attempt to write down the wave function of the universe, we are allowed to quantize the universe using the Wheeler-DeWitt equation.\footnote{This refutes the anti Wheeler-DeWitt equation argument expressed in DeWitt  \cite{DeWitt99, DeWitt03}. The Wheeler-DeWitt equation is the equation for the vacuum state of the universe.} We shall construct a quantized Friedman Robertson Walker universe in which the only field is a gage field (actually a radiation field). Imposing the boundary condition that classical physics hold exactly at ``late times" (any time after the first minute) implies that classical physics is good all the way into the initial singularity. 

 Recall that the Wheeler-DeWitt equation is 

\begin{equation} \hat {H}\Psi=0
\label{eq:WheeDeW} \end{equation}	
 \noindent where $\hat{H}$ is the super-Hamiltonian operator. This operator contains the equivalent of time derivatives in the Schrodinger equation. We say ``the equivalent'' because the Wheeler DeWitt equation does not contain time as an independent variable. Rather, other variables-matter or a spatial metric-are used as time markers. In other words, a variation of the physical quantities {\it is} time. Depending on the variable chosen to measure time, the time interval between the present and the initial or final singularity may be finite or infinite-but this is already familiar from classical general relativity. In the very early universe, conformal time measures the rate at which particles are being created by  tunneling, that is, it measures the rate at which new information is being created. Therefore, the most appropriate physical time variable is conformal time, and thus we shall select an appropriate combination of matter and spatial variables that will in effect result in conformal time being used as the fundamental time parameter in the Wheeler-DeWitt equation.

 Julian Barbour  \cite{10} notes that the Wheeler-DeWitt equation {\it is independent of time}. This shouldn't surprise us, because we know that there is a reference frame in which the universe is timeless. However, Barbour goes so far as to conclude that time that flows might be merely an illusion.\footnote{Brian Greene  \cite{11} points out that Einstein's theory of special relativity {\it requires} that {\it all of space-time}, that is, all of space and all of time, be {\it present at once}.}

The universe pictured here is really the ground state of a universal wave function, and we can use a natural metric on super space given by the dynamics of this ground state, which satisfies the Wheeler DeWitt equation. This natural metric is a two dimensional hyperbolic metric with signature $-+$. We stay within the lowest-order the semi classical cosmology approximation for the Friedman Robertson Walker geometry of radius $a$ coupled to a scalar field $\phi$, and use the  Wheeler-DeWitt equation. The argument we shall give is independent of the dynamics; it only depends on the basic structure of quantum mechanics and Riemann's geometry. A dynamical argument would be sticky if one does not want to make any a priory assumptions about the cosmological constant.

The Hilbert action $S$ in the ADM formalism can be written

\begin{equation} S=\int R\sqrt{-g}\, d^4x=\int L\, d t
\label{eq:ADMaction} \end{equation}	
 \noindent where $R$ is the curvature scalar as before. If matter is in the form of a radiation field (string theories, in particular) the effective action in four dimensions is given by the expression, 
\begin{equation}
\label{lg2}
{\it L}=\sqrt{-g}e^{-\phi}\left(R+\phi_{;\rho}\phi^{;\rho}\right), 
\end{equation} where $\phi$ is the radiation field and it has been set $\hbar=1$. If the space-time is assumed to be the Friedman Robertson Walker universe
\begin{equation} \label{m} d s^2=-N^2 {d}t^2+\frac{{a(t)}^2}{1+\frac{\epsilon}{4}r^2}[{d}r^2+r^2 ({d} \theta ^2+\sin ^2 (\theta) {d} \phi ^2)], 
\end{equation}
containing isotropic radiation (where the spatial curvature $\epsilon$ takes the values 0, $1$, $-1$). In  \cite{Rubakov77}, the authors have shown the canonical variables can be chosen $(a, \phi)$, where $a$ is the scale factor of the universe, and $\phi$ is a particular parametrization of the field variables called {\it Schultz potentials}  \cite{Schutz71}. (We are not using the same symbol for the curvature scalar and the scale factor of the universe.) The momenta conjugate to these canonical variables will be written $(\pi_a, \pi_\phi)$.

In the Lagrangian (\ref{lg2}) we insert the metric (\ref{m}). The ADM Lagrangian in these variables can be shown to take the form

\begin{equation} L=N(H_g +H_m)
\label{eq:ADML} \end{equation}
where
\begin{equation}
H_g=-\frac{\pi^2_a}{2a}-\frac a2
\end{equation}
 \noindent is a purely gravitational super-Hamiltonian, and
 
\begin{equation}
H_m=\frac{a^3\dot\phi^2}{2N^2}=\frac{\pi^2_\phi}{a^3} 
\end{equation}
 \noindent is both a coordinate energy density measured by a co-moving observer and the super-Hamiltonian of the matter.	The dot denotes the time derivative.

The momentum conjugate to $a$, the scale factor of the universe, is
 
 \begin{equation} \pi_a=\frac{a\dot a}{ N}.
 \label{eq:pR} \end{equation}	
 A constraint equation for the Friedman universe is obtained by substituting (\ref{eq:ADML}) through (\ref{eq:pR}) into (\ref{eq:ADMaction}) and varying the lapse $N$. The result is the super-Hamiltonian constraint:

\[
0=H_g +H_m=-\frac{\pi^2_a}{ 2a}-\frac a2 +\frac{\pi^2_\phi}{a^3}.
\]
If we choose the momentum conjugate to the {\it true} time $\tau$ to be the term

\[
\pi_\tau=\frac{\pi^2_\phi}{a^2}
\]
 \noindent then the super-Hamiltonian constraint becomes

\begin{equation} 0=-\frac{\pi_a^2}{ 2a}-\frac a2 +\frac{\pi_\tau}{ a}
 \label{eq:tauH} \end{equation}	
 \noindent which is just the Hamiltonian for a simple harmonic oscillator. If we quantize by the replacement $\pi_\tau \rightarrow \hat \pi_\tau=-i \partial /\partial \tau$, and $\pi_a \rightarrow \hat \pi_a=-i\partial /\partial a$, together with a reversal of the direction of time $\tau\rightarrow-\tau$ in the super-Hamiltonian constraint (\ref{eq:tauH}), the Wheeler-DeWitt equation (\ref{eq:WheeDeW}) will then become (if we ignore factor ordering problems) Schrodinger equation for a simple harmonic oscillator with mass $m=1$, spring constant $k=1$ and angular frequency $\omega=1$:
 
\[
i{\partial\Psi \over\partial \tau}={1\over 2}{\partial^2\Psi\over\partial a^2} -\frac{a^2}2\Psi.
\]
The Grady relations  arose 20 years ago in a very interesting paper \cite{dg}, where  Grady considered a class of systems described by a Hamiltonian given as a linear combination of two operators obeying some nonlinear relations. The relations, afterward called by their names, guarantee the existence of an infinite set of mutually commuting particles which includes the Hamiltonian. This naturally connects the construction to the integrable systems \cite{dg, Perk, dav, gehlen, baxter}. The nonlinear relations can be used to construct the infinite set of mutually commuting operators for some integrable systems.

The infinite set of the mutually commuting operators constructed by  Grady \cite{dg} is represented as\begin{equation}\label{lin} 2J_m=A_m+A_{-1}+a\left(A_{m+1}+A_0\right),\end{equation} where $a\in\mathbb R$.

The Hamiltonian of the system associated with the space-time field of commuting particles (see Equation (\ref{jmn}) for their explicit form) reads as\begin{equation}\label{H} H=\big(f(a)+{{\phi}^2} g(a)\big)-\frac{\frac{\partial^2}{\partial\phi^2}}{h(a)}+\frac{\partial^2}{\partial a^2},\end{equation} where a real parameter $a$ serves as the scale factor of the universe. The operator $H$ is supposed to obey the nonlinear Grady relations \footnote{In principle, one can also deal with a Hermit operator $H$ . Then the Hamiltonian (and the whole set of the integrals (\ref{jmn})) is Hermit if the coupling constant is pure imaginary \cite{nsusy}. However, here we will not discuss such systems.}. So, let us consider a scalar model given by the action\begin{equation}\label{S2} S={}\int\sqrt{-g}e^{-\phi}\left(\phi_{;\rho}\phi^{;\rho} +R\right)d t,\end{equation} where $\phi_{;\rho}$ is a co-variant derivative of a scalar field and $R$ is the curvature scalar. This term can be treated as describing the interaction of the vector field $\phi_{;\rho}$ with an external symmetric field $R$. One can also think that it contains the mass term of the vector field. The scalar models of the form (\ref{S2}) are widely used in the context of string theories.

The simple harmonic oscillator is defined by the algebra generated by the set of operators $\{\hbox{{1\hskip -5.8pt 1}\hskip -3.35pt I},\phi, a\}$ where $a$ is the scale factor of the universe. As a consequence of the natural grading of the  algebra defined by the operator $a$, for this representation of the generating element $H$ it is possible to realize the Grady relations only in its contracted form. The non commutative system admits the following two types of gauge transformations:\begin{align}\label{transV}\Psi'(a,\phi)&=U(a,\phi)\Psi(a,\phi),&
\hskip 2cm\Psi'{}^{*}(a,\phi)&=\Psi^{*}(a,\phi)U^{*}(a,\phi),\end{align} or\begin{equation}\label{transT}\Psi'(a,\phi)=U(a,\phi)\Psi(a,\phi)U^{*}(a,\phi)\end{equation} with $U(a,\phi)$ being a unitary operator. In the commutative limit, the transformation (\ref{transV}) is reduced to the usual $U(1)$ gauge transformation while (\ref{transT}) becomes trivial. Let us suppose that the gravitational field depends on $a$ only. This corresponds to a radially symmetric field in the commutative limit. The following choice of the gauge,\begin{equation}\label{gauge} \hskip 2cm A(a,\phi)=f(a)\phi,\end{equation} where $f( \cdot)$ is a real function, guarantees such a dependence of the gravitational field. In general, the operator $H_0\equiv H$ together with the contracted Grady relations recursively generate the infinite-dimensional contracted Onsager algebra \cite{nsusy}:\begin{align}\label{cOA}
\left[ H_g,\: H_m\right]&=3\,{\frac {\left( {\frac {\partial ^{3}}{\partial\phi\sp2\partial a}} \right) a-2\,{\frac {\partial\sp2}{\partial\phi\sp2}}}{{a}^{6}}},&\left[ H_g,\:B_0\right]&= H_0.\end{align} This algebra can be extended discarding the last condition. The algebra (\ref{cOA}) admits the infinite set of the commuting quadratic particles\begin{equation}\label{jmn}\mathsf J^g=\left\{ H_{g},\, H_0\right\} -3\left((k-1)B_0-\frac a 2\right)\,{\frac {\left( {\frac {\partial ^{3}}{\partial\phi\sp2\partial a}} \right) a-2\,{\frac {\partial\sp2}{\partial\phi\sp2}}}{{a}^{6}}}\end{equation} which contains the Hamiltonian (\ref H), and the grading operator,
\[
\mathsf J^0= 3\,{\frac {\left( {\frac {\partial ^{3}}{\partial\phi\sp2\partial a}} \right) a-2\,{\frac {\partial\sp2}{\partial\phi\sp2}}}{{a}^{6}}}.
\]
We introduce a vector (``ground'' state)  which obeys the conditions\begin{align}\label{main} H\Psi&={\mathcal D}\alpha\Psi^{*},& H\Psi^{*}&={\mathcal D}\beta\Psi,& S\Psi^{*}&=s\Psi^{*}.\end{align} Here $\alpha(\cdot)$ and $\beta(\cdot)$ are some functions, and $s\in\mathbb R$. We would like to note that although we use formally the  conditions (\ref{main}), one should remember that a potential disadvantage of the operator  technique is that  the ordering of the operators may not be commutative.
\subsection{Solution for a mass-less scalar field} For the case when the scalar field is mass-less, the Wheeler-DeWitt equation in the mini super space reads as (\ref{ww2}) with
\begin{eqnarray}
f(a)&=&-a^2,\nonumber\\
g(a)&=&0,\nonumber\\
\label{}
h(a)&=&a^2/2.
\end{eqnarray}
We use the separation of variables method writing $\Psi(a, \phi)$ as a product of
\begin{itemize}
\item a function of the scaling parameter, and
\item a function of the field.
\end{itemize}
The solution of (\ref{ww2}), is
\begin{eqnarray} \Psi(a, \phi)&=&a^{1/2}\left[AI_n\left(-\frac{a^2}4\right)+B\right]\times\nonumber\\
&&\phi^{1/2}\left[CI_m\left(\phi^2\right)+D K_m\left(\phi^2\right)\right],
\end{eqnarray}
where $I_n$ and $K_m$ denote the modified Bessel functions, and the coefficients $A$, $B$, $C$ and $D$ are constant.

In general, the solution may be an exponentially growing or decreasing function of $a$. If the order of the modified Bessel functions is positive, the solution may exhibit an oscillatory qualitative behavior. However for this case, the function oscillates for small values of $a$, increasing or decreasing for large values of $a$, suggesting that a classical phase may occur only for small $a$.
\subsection{A massive scalar field minimally coupled to gravity}
 A particle is a point field in space-time characterized by a Hilbert action. The resulting equations of motion couple that point with the space-time field with a factor of $m$/$m_{Pl}$, where $m$ is the mass of the particle. Thus, the gravity component of the action is very weakly coupled to particles.\footnote{The electromagnetic component of an action is also affected by this coupling.} The difference between the mass-less and the massive case is given by the formulas (\ref{}). Indeed, the fact that the mass of the scalar field is damped by a Planck mass is not equivalent to saying that  mass-less and massive scalar fields are equivalent.

We solve the Wheeler-DeWitt equation equation for a massive scalar field. 

We consider three smooth functions $f$, $g$ and $h$. We generalize (\ref{eq:ADML}) as
\begin{equation} \label{lq} L=\frac{a{\dot a}^2}{2N}-h(a)\frac{a{\dot\phi}^2}{2N}-\frac{\phi^2g(a)N}{2a}-\frac{f(a)N}{2a}.\end{equation} \par From (\ref{lq}) we obtain the conjugate momenta, \begin{eqnarray}
\label{10}
\pi_a&=&\frac{a\dot a}{N}, \\
\label{11}
\pi_\phi&=&-h(a)\frac{a\dot\phi}{N}.\end{eqnarray} We construct the Hamiltonian $H$, which takes the form 
\[
H=N\left[\frac{\pi_a^2}{2a}-\frac{\pi_\phi^2} {2h(a)a}-\frac{\phi^2g(a)}{2a}-\frac{f(a)}{2a}\right].
\]
Variation of $N$ yields the first class constraint ${H} \approx0$. The Dirac quantization procedure yields the Wheeler-DeWitt equation by imposing the condition (\ref{eq:WheeDeW}) and performing the substitutions\begin{eqnarray*} \pi_a^2&\rightarrow&-\frac{\partial^2}{\partial a^2}, \\ \pi_\phi^2&\rightarrow&-\frac{\partial^2}{\partial\phi^2}.\end{eqnarray*}
The Wheeler-DeWitt equation in the mini super space reads 
\begin{equation}\label{ww2}
\big(f(a)+{{\phi}^2} g(a)\big) \Psi (a, \phi)-\frac{\frac{\partial^2\Psi}{\partial\phi^2}(a, \phi)}{h(a)}+\frac{\partial^2\Psi}{\partial a^2}(a, \phi)=0.
\end{equation}
Discarding the Zero field, we solve equation (\ref{ww2}). We explain of the adopted procedure by justifying the algorithm used for finding solutions avoiding the separation of the variables. Suppose that:
\begin{itemize}
\item the second partial derivative of $ \ Psi $ with respect to the variable $ a $ is zero,
\[
\frac{\partial^2\Psi}{\partial a^2}(a, \phi)=0,
\]
and
\item  $f$ is zero.
\end{itemize}
So Bessel functions solve the equation (\ref{ww2}). Not supposing separability of this equation, we propose a power series of the field with coefficients that depend on the radius,
\begin{equation}
\label{1}
\sum_{n=0}^\infty{{\phi }^{\frac{1}2+2 n}} {{\psi }_n}(a).
\end{equation}
We give a tentative to obtain additional information avoiding the separation of the variables in choice (\ref1). By the coefficient of ${\phi }^{\frac{1}2+2 n}$ in Equation (\ref{ww2}), we obtain the equations for these coefficients leading to infinite ordinary differential equations for $\psi_1$, $\psi_2$, ...
\[
4 g {{\psi}_{-1+n}}+4 f {{\psi}_n}+\frac{(-3-4 n) (5+4 n) {{\psi}_{1+n}}}{h}+4 \psi _{n}^{''}=0,
\]
with $\psi_0$ being an integration function.
For each $n$, we define the functions
\begin{eqnarray*}
{{\alpha}_n}&=&\frac{4 g h}{(3+4 n) (5+4 n)}, \\
{{\beta}_n}&=&\frac{4 f h}{(3+4 n) (5+4 n)}, \\
{{\gamma}_n}&=&\frac{4 h}{(3+4 n) (5+4 n)}, \\
\end{eqnarray*}
then,
\begin{equation}
\label{3}
{{\psi}_{1+n}}={{\alpha}_n} {{\psi}_{-1+n}}+{{\beta}_n} {{\psi}_n}+{{\gamma}_n} \psi _{n}^{''}.
\end{equation}
For each $m$ and each $n$ natural numbers, we define the functions
\[
{f_{0, 0}}=1,
\]
\[
{f_{2m, 1+n}}={{\alpha}_n} {f_{2 m,-1+n}}+{{\beta}_n} {f_{2 m, n}}+{{\gamma}_n} ({f_{2 (-1+m), n}}+4 f_{-1+2 m, n}^{\prime}+f_{2 m, n}^{''}),
\]
\[
{f_{1+2m, 1+n}}={{\alpha}_n} {f_{1+2 m,-1+n}}+{{\beta}_n} {f_{1+2 m, n}}+{{\gamma}_n} ({f_{-1+2 m, n}}+f_{2 m, n}^{\prime}+f_{1+2 m, n}^{''}).
\]
Let us replace (\ref {3}) in (\ref {1}). 
\[
\sum_{n=0}^\infty{{\phi }^{\frac{1}2+2 n}} {{\psi }_n}
\]
\[
=\sqrt\phi\left({{\psi }_0} \big(1+{{\phi }^2} {{\beta }_0}+{{\phi }^4} ({{\alpha }_1}+{{\beta }_0} {{\beta }_1}+{{\gamma }_1} \beta _{0}^{'' })\big)+2 {{\phi }^4} {{\gamma }_1} \beta _{0}^{\prime } \psi _{0}^{\prime }\right.
\]
\[
\left.+\big({{\phi }^2} {{\gamma }_0}+{{\phi }^4} ({{\beta }_1} {{\gamma }_0}+{{\beta }_0} {{\gamma }_1}+{{\gamma }_1} \gamma _{0}^{'' })\big) \psi _{0}^{'' }+2 {{\phi }^4} {{\gamma }_1} \gamma _{0}^{\prime } \psi _{0}^{(3)}+{{\phi }^4} {{\gamma }_0} {{\gamma }_1} \psi _{0}^{(4)}+...\right)
\]
\begin{equation}
\label{4}
={\sqrt{\phi }}\hspace{1em}\sum _{n=0}^{\infty }{{\phi }^{2n}}\sum _{m=0}^{\infty }\big({f_{2m,n}}\psi _{0}^{(2m)}+2{f_{1+2m,n}}\psi _{0}^{(1+2m)}\big),
\end{equation}
which is the complete general solution $\Psi$ of the Wheeler-DeWitt equation.

If $ \psi_0$ has compact support, then (\ref {4}) has compact support in $a$. The function does exhibit an oscillatory behavior if $\psi_0$ oscillates. For this case, if the function $\psi_0$ oscillates for small values of $a$, and increases or decreases for large values of $a$, then a classical phase may occur for small values of $a$ only. The function $\Psi$ has this oscillatory behavior for all $\phi$.

We think that this procedure can be developed more deeply to obtain a robust method to obtain other solutions for the Wheeler-DeWitt equation.

\begin{figure}
\psfig{file=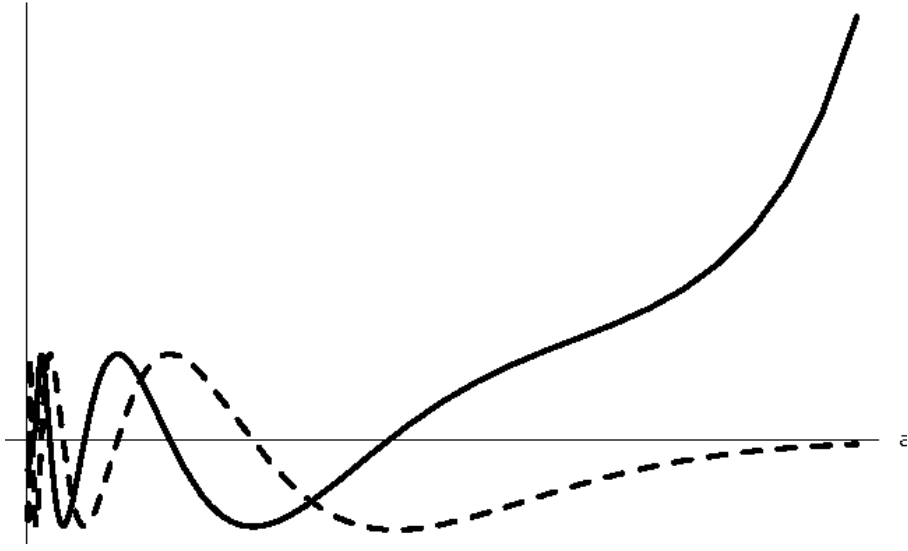,width=12cm}
\caption{Qualitative behavior of $\Psi(a, \phi)$ for the massive scalar field case when $\psi_0$ oscillates with $\psi_0=\exp(\pm a)\cos a^{-1}$. The dashed and continuous lines represent the positive and negative exponents of $\psi_0$ respectively.}
\end{figure}
\subsection{Kramers approximation for a scalar field}
One way to obtain the transition to the classical regime from the quantum solution might be to employ a Kramers approximation, like in usual quantum mechanics. This can be achieved by rewriting the wave function as, 
\[
\Psi=\exp\left({\frac{i}{\hbar}S}\right), 
\]
substituting it into the Wheeler-DeWitt equation, and performing an expansion in orders of $\hbar$ in $S$,
\[
S=S_0+\hbar S_1+\hbar^2S_2+...
\]
A classical solution has to be recovered by constructing a wave packet from $S_0$: 
\[
\Psi=\int{\psi_0(k_0)\exp{\left(\frac{i}{\hbar}S_0\right)d k_0}}, 
\]
where $k_0$ is an integration variable. We will analyze a Kramers approximation for a scalar field. Regarding  the problem of ordering in the conjugate momenta,  one can avoid the ordering problem at the semi classical level, like the subject of this paragraph. In this case, it is $S=S(a, \phi)$, and the Kramers expansion in the mini super space Wheeler-DeWitt equation leads to the following equations connecting $S_0$ and $S_1$:
\[
\left(\frac{\partial S_0}{\partial a}\right)^2-\frac1{h(a)}\left(\frac{\partial S_0}{\partial\phi}\right)^2-f(a)=0;
\]
\[
 \left[i\left(\frac{\partial^2S_0}{\partial a^2}\right)-2\left(\frac{\partial S_0}{\partial a}\right)\left(\frac{\partial S_1}{ \partial a}\right)\right]-\frac1{h(a)}\left[i\left(\frac{\partial^2S_0}{\partial\phi^2}\right)-2\left(\frac{\partial S_0}{\partial\phi}\right)\left(\frac{\partial S_1}{\partial\phi}\right)\right]=0.
\]
First we get a solution for $S_0$. The solution can be obtained by taking, 
\[
S_0(a, \phi)=S_0(a)+S_0(\phi), 
\]
leading to two differential equations:\begin{eqnarray*}\left(\frac{{d} S_0(a)}{{d} a}\right)^2&=&\frac{k_0}{h(a)}+f(a), \\\left(\frac{{d} S_0(\phi)}{{d}\phi}\right)^2&=&k_0, \end{eqnarray*} where $k_0$ is a separation constant. These equations admit the following analytic solution:\begin{eqnarray} \label{bri} S_0(a)&=&{A_0}\pm\int _{0}^{a}{\sqrt{f(x)+\frac{{k_0}}{h(x)}}}{d x}, \\
S_0(\phi)&=&{B_0}\pm\phi {\sqrt{{k_0}}},\nonumber
\end{eqnarray} where $A_0$ and $B_0$ are integration constants. We follow the same procedure in order to obtain a solution for $S_1(a, \phi)$, considering first $S_1(a, \phi)=S_1(a)+S_1(\phi)$. We get the solution, \begin{eqnarray*} S_1(a)&=&{A_1}+\frac{1}{4} \int _{0}^{a}\frac{\pm2 {\sqrt{f(x)+\frac{{k_0}}{h(x)}}} {k_1}+i h(x) {f^{\prime }}(x)-\frac{i {k_0} {h^{\prime }}(x)}{h(x)}}{f(x) h(x)+{k_0}}{d x},\\ S_1(\phi)&=&{B_1}\pm\frac{\phi {k_1}}{2 {\sqrt{{k_0}}}},\end{eqnarray*} where $A_1$ and $B_1$ are integration constants. From the solution for $S_0(a)$, we easily see that only when $k_0>0$ we obtain an oscillatory qualitative behavior of the wave function for small values of $a$, while when $k_0 < 0$ the wave function has an exponential qualitative behavior for any value of $a$. Similarly, if $k_0>0$, then $\exp{[\frac{i}{\hbar}S_0(\phi)]}$ is oscillatory for any value of $\phi$, otherwise the solution has an exponential qualitative behavior. Hence, for $k_0>0$, $\exp{\frac{i}{\hbar}S_0(a, \phi)}$ oscillates for small values of $a$ and any value of $\phi$.

 We construct a wave packet from the solution through the expression, 
\begin{equation} \label{wp} \Psi(a, \phi)=\int{\psi_0(k_0)\exp{\left[\frac{i}{\hbar}S_0(k_0, a, \phi)\right]d k_0}},\end{equation} where the function $\psi_0(k_0)$ may be a sharply peaked Gaussian centered in $\bar k_0$, with width $\sigma$. Examining Equation (\ref{bri}), we see that $S_0 (a)$ becomes very large when $a$ becomes very small, and $f$ and $h$ are of quadratic order. Hence, in the integral (\ref{wp}), constructive interference happens only if 
\[
\frac{\partial S_0(a, \phi)}{\partial k_0}=0, 
\]
which implies a relation between $k_0, a$ and $\phi$, $k_0=k_0(a, \phi)$. The wave function turns out to be: 
\begin{equation} \label{wp2} \Psi(a, \phi)=\psi_0[k_0(a, \phi)]\exp\left\{\frac{i}{\hbar}S_0[k_0(a, \phi), a, \phi]\right\}.\end{equation} As the Gaussian is sharply peaked at $k_0(a, \phi)=\bar k_0$, then we obtain that the wave function (\ref{wp2}) is sharply peaked at $\bar k_0$.
\subsection{Complex time in the Wheeler-DeWitt formalism}
\label{s:KGcomplex}We    discuss an unresolved problem of the Wheeler-DeWitt formalism,  the problem of positive-definite {\em physical} inner product in the space of physical solutions of this formalism.

An imaginary contribution to time can be seen also from the well-known physical inner product formulas available for the Wheeler-DeWitt formalism. First, we consider the free relativistic particle in $1+1$ dimensions, described by a complex-valued scalar wave function of two variables, $\psi(x_0, x_1)$, subject to the constraint \begin{equation} \left( -\hbar^2 \frac{\partial^2}{\partial x_0^2} + \hbar^2 \frac{\partial^2}{\partial x_1^2} - m^2 \right) \psi(x_0, x_1) = 0 \,. \label{eq:free_rel_PDE} \end{equation}General solutions have the form \begin{equation} \psi_{\rm phys} (x_0, x_1) = \int_{-\infty}^{\infty} \left( f_+(k) e^{i\hbar^{-1}(k x_1 - \epsilon_k x_0)} +  f_-(k) e^{i\hbar^{-1}(k x_1 + \epsilon_k x_0)} \right) {\rm d}x_1 \,, \label{sol_KG_PDE} \end{equation} where $\epsilon_k = \sqrt{k^2+m^2}$. Solutions in this general form automatically split into positive frequency and negative frequency components, a split which is important for constructing the physical Hilbert space (see ,e.g., \cite{GenRepIn}). On positive frequency solutions, the physical inner product is \begin{equation} \left( \phi, \psi \right) = \left. i\hbar \int_{-\infty}^{\infty} \left(\bar{\phi}(x_0, x_1) \frac{\partial}{\partial x_0} \psi(x_0, x_1) -  \psi(x_0, x_1) \frac{\partial}{\partial x_0} \bar{\phi}(x_0, x_1)  \right) {\rm d}x_1 \right|_{x_0=t} \label{eq:KG_prod} \end{equation} with an extra minus sign for negative frequency solutions, while negative frequency and positive frequency solutions are mutually orthogonal. When evaluated on solutions to (\ref{eq:free_rel_PDE}), the integration is independent of the value of $t$. As a simple (deparametrizable) example consider once again the free relativistic particle subject to the constraint equation~(\ref{eq:free_rel_PDE}). This equation is hyperbolic and the initial value problem is a priory well-posed, but a general solution~(\ref{sol_KG_PDE}) will include both positive and negative frequencies. Consequently, the constant-time inner product given by~(\ref{eq:KG_prod}) fails to be positive-definite and cannot on its own provide us with a physically meaningful unitary interpretation of the evolution. Only if we impose the further restriction of only considering, e.g., positive frequency modes, do we have a positive-definite physical inner product and a physically meaningful solution to the initial value problem. The latter is owed to the fact that restriction to positive frequencies is tantamount to imposing a (in this case forward pointing) time direction.\footnote{Also in the classical treatment of relativistic systems, where the square of the momentum conjugate to the clock appears in the constraint, one is required to specify the time direction in order to formulate a relational initial value problem. Namely, given the initial data of the other variables at the initial value of the clock, one can only solve the constraint up to sign for the momentum conjugate to the clock. One is forced to choose a sign which then determines the time direction. } It seems hardly imaginable that, in more general scenarios with frequency mixings, inner products relying on constant clock-time surfaces are meaningful. These are usually also closely linked to an --- at least local --- unitary evolution of initial data in some clock time, generated by some suitable Hamiltonian. But in a highly quantum state of a system with no global time even local unitary evolution becomes meaningless close to the turning region where frequency mixing is significant --- apart from the fact that positive and negative frequencies require two separate Hamiltonians for evolution. A physical inner product based on more general boundaries or on the entire configuration space is in general required to cope with such highly quantum scenarios.
\section{The Perspective of the Causal Interpretation}
The Hamiltonian can be reduced to general mini super space form: 
\[
H_{GR}=N(t) {H}(p^{\alpha}(t), q_{\alpha}(t)), 
\]
where $p^{\alpha}(t)$ and $q_{\alpha}(t)$ represent the homogeneous degrees of freedom coming from $\Pi ^{i j}(x, t)$ and $h_{i j}(x, t)$. The mini super space Wheeler-De Witt equation is: 
\begin{equation}
\label{bsc} {H}({\hat{p}}^{\alpha}(t), \hat q_{\alpha}(t)) \Psi (q)=0.\end{equation} Writing $\Psi=R\exp (i S/\hbar)$, and substituting it into (\ref{bsc}), we obtain the following equation: 
\begin{equation} \label{hoqg} \frac{1}2f_{\alpha\beta}(q_{\mu})\frac{\partial S}{\partial q_{\alpha}} \frac{\partial S}{\partial q_{\beta}}+U(q_{\mu})+Q(q_{\mu})=0, 
\end{equation} where 
\begin{equation} \label{hqgqp} Q(q_{\mu})=-\frac{1}{R} f_{\alpha\beta}\frac{\partial ^2 R} {\partial q_{\alpha} \partial q_{\beta}}, 
\end{equation} and $f_{\alpha\beta}(q_{\mu})$ and $U(q_{\mu})$ are the mini super space particularizations of the DeWitt metric $G_{i j k l}$ \cite{dew} and the scalar curvature density $-h^{1/2}R^{(3)}(h_{i j})$ of the space like hyper surfaces. The causal interpretation, applied to quantum cosmology, states that the trajectories $q_{\alpha}(t)$ are real, independently of any observations. Equation (\ref{hoqg}) is the Hamilton-Jacobi equation for them, which can be a classical one amended with a quantum potential term (\ref{hqgqp}), responsible for the quantum effects. This suggests to define: 
\[
p^{\alpha}=\frac{\partial S}{\partial q_{\alpha}}, 
\]
where the momenta are related to the velocities in the usual way: 
\[
p^{\alpha}=f^{\alpha\beta}\frac{1}{N}\frac{\partial q_{\beta}}{\partial t} \;.
\]
To obtain the quantum trajectories we solve the following system of first order differential equations: 
\begin{equation} \label{h3} \frac{\partial S(q_{\alpha})}{\partial q_{\alpha}}=f^{\alpha\beta}\frac{1}{N}\frac{\partial q_{\beta}}{\partial t} \;.\end{equation} Equations (\ref{h3}) are invariant under time re-parametrization. Hence, even at the quantum level, different choices of $N(t)$ yield the same space-time geometry for a given non-classical solution $q_{\alpha}(t)$. There can not be problem of time in the causal interpretation of mini super space quantum cosmology. Let us then apply this interpretation to our mini super space models and choose the gage $N=1$.
\subsection{The Born Interpretation}
Since $\rho(a(t))\equiv\Psi \Psi^\ast={ R}^2$ measures the density of universes with radius $a$, for normalizable wave functions, it implies the Born Interpretation: the probability that we will find ourselves in a universe with size $a$ is proportional to ${ R}^2$. Similarly, if ${ R}=\rm\, constant$, we are equally likely to find ourselves in a universe with any given radius. However, since $a>0$, if we ask for the relative probability that we will find ourselves in a universe with radius larger than any given radius $a_{given}$ or instead find ourselves in a universe with radius smaller than $a_{given}$, we see that the relative probability is one that we will find ourselves in a universe with radius larger than $a_{given}$, since $\int^\infty_{a_{given}}{ R}^2\, d { R}=\infty$ while $\int^{a_{given}}_0{ R}^2\, d { R}$ is finite. Thus with probability one we should expect to find ourselves in a universe which if closed is nevertheless arbitrarily close to being flat. This resolves the Flatness Problem in cosmology, and we see that we live in a near flat universe because (1) the Copenhagen Interpretation applies in the large, or equivalently, because (2) a quantum universe began as a delta function at the initial singularity, or equivalently, because (3) classical physics applies on macroscopic scales. Notice a remarkable fact: although the calculation was done using the Wheeler-DeWitt equation, the same result would have been obtained if we had done it in classical GR (in its Hamilton-Jacobi form), or even done it in Newtonian gravity (in its Hamilton-Jacobi form). Just as one can do Friedman Robertson Walker cosmology in Newtonian gravity, so one may also do Friedman Robertson Walker cosmology in quantum gravity. The conclusion is the same in all theories: the universe has to be flat. This conclusion does not, in other words, depend on the value of the speed of light, or on the value of Planck's constant. In short, the flatness conclusion is robust!
	
\section{Conclusion}
We have calculated the Bohemian trajectories for exact wave solutions of the Wheeler-DeWitt equation. All of them present the same qualitative behavior.

After quantizing the models, we have obtained the general solution of the Wheeler-De Witt equation. Usually, the solutions are oscillatory when the scale factor is small and not oscillatory when the scale factor becomes large. This suggests that non-classical qualitative behavior may occur when the scale factor is large. \par We conclude this paper by stating that in quantum cosmology we have not necessary that the classical qualitative behavior appears when $a$ is large, while quantum qualitative behavior may be present when $a$ is small. It can indeed be the reverse. This was already pointed out in \cite{glik} and we presented specific examples illustrating this fact.
\end{document}